\documentclass[preprint,aps,showpacs,superscriptaddress]{revtex4-1}
\usepackage{graphicx}
\usepackage{dcolumn}
\usepackage{amsfonts, amssymb}

\def\bea{\begin{eqnarray}}
\def\eea{\end{eqnarray}}
\def\beq{\begin{equation}}
\def\eeq{\end{equation}}

\newcommand{\kkbar}{\bar{K}{K}}
\newcommand{\kbarkbar}{\bar{K}\bar{K}}

\begin{document}

\title{Exact calculations of a quasi-bound state in the $\bar{K} \bar{K} N$ system}

\author{N.V. Shevchenko\footnote{Corresponding author:
shevchenko@ujf.cas.cz}}

\affiliation{Nuclear Physics Institute, 25068 \v{R}e\v{z}, Czech Republic}

\author{J. Haidenbauer}

\affiliation{Institute for Advanced Simulation and J\"ulich Center for Hadron
Physics, Forschungszentrum J\"ulich, D-52425 J\"ulich, Germany}

\date{\today}

\begin{abstract}
Dynamically exact calculations of a quasi-bound state in the $\bar{K}\bar{K}N$
three-body system are performed using Faddeev-type AGS equations. 
As input two phenomenological and one chirally motivated
$\bar{K}N$ potentials are used, which describe the experimental information on
the $\bar{K}N$ system equally well and produce either a one- or two-pole
structure of the $\Lambda(1405)$ resonance. 
For the $\bar{K}\bar{K}$ interaction separable potentials are employed that
are fitted to phase shifts obtained from two theoretical models.
The first one is a phenomenological $\bar{K}\bar{K}$ potential based on
meson exchange, which is derived by SU(3) symmetry arguments
from the J\"ulich $\pi \pi - \bar{K} K$ coupled-channels model.
The other interaction is a variant of the first one,
which is adjusted to the $KK$ s-wave scattering length
recently determined in lattice QCD simulations.
The position and width of the $\bar{K}\bar{K}N$ quasi-bound state is evaluated
in two ways: 
(i) by a direct pole search in the complex energy plane and 
(ii) using an ''inverse determinant'' method, 
where one needs to calculate the determinant of the AGS system of equations
only for real energies. A quasi-bound state is found with binding energy
$B_{\bar{K}\bar{K}N} = 12 - 26$ MeV and width $\Gamma_{\bar{K}\bar{K}N} = 61 - 102$ MeV,
which could correspond to the experimentally observed $\Xi(1950)$ state.
\end{abstract}

\maketitle

\section{Introduction}
\label{Introduction_sect}

It is generally accepted that the $\bar{K}N$ interaction in the isospin-zero state is 
strongly attractive and produces a quasi-bound state, which shows itself as the
$\Lambda(1405)$ resonance in the lower $\pi \Sigma$ channel. This fact
led to the conjecture that a quasi-bound state could also exist in the 
$\bar{K}NN$ three-body system. The first paper on the topic appeared 
more than 50 years ago~\cite{Nogami}, but it was Ref.~\cite{YA} which arouse a large interest
in the question of the existence and properties of such a state. It triggered
many theoretical papers devoted to the $K^- pp$ system and in the sequel
also experimental efforts. However, so far the experimental results neither agree 
with theoretical predictions nor between themselves.
Indeed, the only point on which all theorists agree, is that a quasi-bound state
in the $K^- pp$ really exists. But the actual predictions for its binding energy 
and width vary over a fairly wide range.

A study with the aim to find a similar quasi-bound state in the $K^- d$ system, which
differs from the $K^- pp$ by quantum numbers ($J^P=1^-$ versus $0^-$)
gave negative results~\cite{PRC_I}.

Another possible candidate for a three-body system with a quasi-bound state is $\bar{K}\bar{K}N$.
In contrast to $\bar{K}NN$, however, it contains the $\bar{K}\bar{K}$ interaction, 
which is expected to be repulsive \cite{Griffith:1968,Chen:2006}. Thus, the principal 
question is whether the repulsion is strong enough to exclude the possibility of a 
quasi-bound state formation in the system. A first exploratory investigation of the
$\bar{K}\bar{K}N$ system was presented in Ref.~\cite{KanadaEn'yo:2008wm}. In this
variational calculation based on a simplified two-body input indeed a bound state
was found. 

In the present paper we report on the first dynamically exact
calculation of a quasi-bound state in the $\bar{K}\bar{K}N$ system.
We solve the Faddeev-type Alt-Grassberger-Sandhas (AGS) equations with
two phenomenological and one chirally motivated $\bar{K}N$ potentials.
The potentials describe experimental data equally well and produce a one-
or two-pole structure of the $\Lambda(1405)$ resonance. 
They were originally designed to address the influence of the number of poles 
of the resonance on various properties of the $\bar{K}N$ and $\bar{K}NN$ systems
(quasi-bound states in $K^- pp$ and $K^- d$, $K^- d$ scattering and
$1s$ level shift and width of kaonic deuterium), an issue which was of relevance
for the strangeness physics community, see
e.g.~\cite{gad,our_first_KNpiSig,my_first_Kd}.

There is no experimental information on the $\kbarkbar$ system. However,
recently lattice QCD calculations of the $KK$ $s$-wave scattering length have become 
available \cite{Sasaki:2013vxa,Beane:2007uh} and we take those results as 
guideline for constructing the $\kbarkbar$ potential. The published scattering 
lengths are small and negative and, thus, suggest a weakly repulsive interaction.
On the phenomenological level one can exploit the fact that the $\kbarkbar$ (or $KK$) 
interaction is related to the one in the $\kkbar$ system via SU(3) flavor symmetry. 
For the latter several potential models can be found in the literature. 
We adopt here the $\pi\pi - \kkbar$ coupled channels model developed by the
J\"ulich group \cite{Lohse90,Janssen95} as starting point for generating another
$\kbarkbar$  interaction. This model, derived in the meson-exchange framework,
predicts a somewhat more repulsive $\kbarkbar$ interaction. In the actual calculation
separable representations are employed, and we construct the $\bar{K}\bar{K}$
potentials in such away that they reproduce the phase shifts of the J\"ulich
$\kbarkbar$  meson exchange model and the scattering length of the lattice QCD
calculation~\cite{Beane:2007uh}, respectively.

The position and width of the $\bar{K}\bar{K}N$ quasi-bound state is evaluated
in two different ways.
First, we perform a direct search for the three-body pole in the complex energy
plane. To do this we find the solvability condition of the homogeneous AGS equations.
In addition we employ the ''inverse determinant'' method proposed and
successfully used in Ref.~\cite{PRC_II} for the $K^- pp$ system. It consists
in the calculation of the inverse determinant of the kernel of the AGS equations
and fitting a Breit-Wigner type function to the peak in its modulus squared.

The paper is structured in the following way: 
The three-body equations, which were derived and solved, are described
in the next section. The two-body interactions, being an input for the
AGS equations, can be found in Section~\ref{Interactions_sect}. The obtained
results are presented and discussed in the Section IV, while the
last section contains our conclusions.

\section{AGS equations}
\label{AGS_sect}

The $\bar{K}N$ system is coupled to other channels, in particular, to 
the $\pi \Sigma$ system, where the $\Lambda(1405)$ resonance is observed.
Since it was shown in~\cite{our_KNN_PRC} that a proper inclusion of the $\pi \Sigma$
channel is important for the $\bar{K}NN$ system, we assume that the same
is also the case for $\bar{K}\bar{K}N$. The three-body Faddeev-type
AGS equations for the $\bar{K} \bar{K} N$ system with coupled 
$\bar{K} \pi \Sigma$ channel are of the form
\begin{equation}
\nonumber
U_{ij}^{\alpha \beta} = (1-\delta_{ij}) \delta_{\alpha \beta} (G_0^{\alpha} )^{-1} 
 + \sum_{k=1}^3 \sum_{\gamma=1}^3
(1-\delta_{ik}) \, T^{\alpha \gamma} _k \, G^{\gamma} _0 \, 
U^{\gamma \beta} _{kj}, \\
\label{AGS}
\end{equation}
where the unknown operators $U_{ij}^{\alpha \beta}$ describe the elastic and
re-arrangement processes 
$j^{\beta} + (k^{\beta}i^{\beta}) \to i^{\alpha} + (j^{\alpha}k^{\alpha})$.
The operator $G_0^{\alpha}$ in Eq.~(\ref{AGS}), which is diagonal in the 'particle'
indices, is the free three-body Green's function.
The Faddeev partition indices $i,j = 1,2,3$ denote simultaneously an interacting
pair and a spectator particle, while the 'particle' indices $\alpha, \beta = 1,2,3$
denote the three-body channels. The partition and 'particle' channel indices denoting
the two-body subsystems are summarized in Table \ref{channels.tab}. 
\begin{table}[h]
\caption{Partition ($i$) and `particle' channel ($\alpha$) indices
of the operators in the AGS system of equations~(\ref{AGS}) denoting the two-body
subsystems. The interactions are further
labelled by the two-body isospin values, entering the system before
symmetrisation.}
\label{channels.tab}
\vskip 0.5cm
\begin{tabular}{cccc}
\hline \hline
\quad $i$ $\setminus$ $\alpha$ \, & \, $1$ ($\bar{K}\bar{K}N$)\, & \, $2$
($\bar{K} \pi \Sigma$)\, & \, $3$ ($\pi \bar{K} \Sigma$)\, \\[\smallskipamount]
\hline
\, 1 \, & $\bar{K}N_{\, I=0,1}$ & $\pi \Sigma_{\, I=0,1}$
    & $\bar{K} \Sigma_{\, I=\frac{1}{2},\frac{3}{2}}$ \\
\, 2 \, & $\bar{K}N_{\, I=0,1}$ & $\bar{K} \Sigma_{\, I=\frac{1}{2},\frac{3}{2}}$
    & $\pi \Sigma_{\, I=0,1}$\\
\, 3 \, & $\bar{K} \bar{K}_{\, I=0,1}$ & $\bar{K} \pi_{\, I=\frac{1}{2},\frac{3}{2}}$
    & $\bar{K} \pi_{\, I=\frac{1}{2},\frac{3}{2}}$ \\[\smallskipamount]
\hline \hline
\end{tabular}
\end{table}

The operators $T^{\alpha \beta}_i$ in Eq.~(\ref{AGS}) are two-body
$T$-matrices immersed into the three-body space. For the description
of the two-body interactions we use separable potentials
\begin{equation}
\label{Voperator}
 V_{i,I}^{\alpha \beta} = \lambda_{i,I}^{\alpha \beta} \,
 |g_{i,I}^{\alpha} \rangle  \langle g_{i,I}^{\beta} | \,,
\end{equation}
which depend on the two-body isospin $I$ and lead to a separable form of the 
corresponding $T$-matrices:
\begin{equation}
\label{Toperator}
 T_{i,I}^{\alpha \beta} = |g_{i,I}^{\alpha} \rangle
\tau_{i,I}^{\alpha \beta} \langle g_{i,I}^{\beta} | \,.
\end{equation}
Separable potentials allow one to introduce new transition and kernel operators
defined via 
\begin{eqnarray}
\label{X_definition}
X_{ij, I_i I_j}^{\alpha \beta} &=&  \langle
g_{i,I_i}^{\alpha} | G_0^{\alpha} \, U_{ij, I_i I_j}^{\alpha
\beta} G_0^{\beta} | g_{j,I_j}^{\beta} \rangle \,, \\
\label{Z_definition}
Z_{ij, I_i I_j}^{\alpha \beta} &=&
\delta_{\alpha \beta} \, Z_{ij, I_i I_j}^{\alpha} =
\delta_{\alpha \beta} \, (1-\delta_{ij}) \,
\langle g_{i,I_i}^{\alpha} | G_0^{\alpha} | g_{j,I_j}^{\alpha}
\rangle \, , 
\end{eqnarray}
to obtain a simpler system of equations than that in Eq.~(\ref{AGS}):
\begin{equation}
\label{full_oper_eq}
X_{ij, I_i I_j}^{\alpha \beta} = \delta_{\alpha \beta} \,
Z_{ij, I_i I_j}^{\alpha} +
\sum_{k=1}^3 \sum_{\gamma=1}^3 \sum_{I_k}
Z_{ik, I_i I_k}^{\alpha} \, \tau_{k, I_k}^{\alpha \gamma} \,
X_{kj, I_k I_j}^{\gamma \beta} \,.
\end{equation}

The free three-body Hamiltonian of the channel $\alpha$ is defined in
momentum representation by
\begin{equation}
\label{H0}
H_0^{\alpha} = \frac{(k_{i}^{\alpha})^2}{2 \, m_{jk}^{\alpha}} +
\frac{(p_{i}^{\alpha})^2}{2 \, \mu_{i}^{\alpha}} \,,
\end{equation}
where $m_{jk}^{\alpha}$ and $\mu_{i}^{\alpha}$ are two- and
three-body reduced masses, respectively
\begin{equation}
m_{jk}^{\alpha} = \frac{m_j^{\alpha} m_k^{\alpha}}
{m_j^{\alpha} + m_k^{\alpha}}, \quad
\mu_{i}^{\alpha} = \frac{m_i^{\alpha} (m_j^{\alpha} + m_k^{\alpha})}
{m_i^{\alpha} + m_j^{\alpha} + m_k^{\alpha}}, \quad i \neq j \neq k \,.
\end{equation}
Three Jacobi momentum coordinate sets $| \vec{k_{i}}^{\alpha},
\vec{p_i}^{\alpha} \rangle$, $i=1,2,3$, $\alpha=1,2,3$ are used, were
$\vec{k_{i}}^{\alpha}$ is the center-of-mass momentum of the $(jk)$ pair
and $\vec{p_{i}}^{\alpha}$ is the momentum of the spectator $i$ with respect
to the pair $(jk)$, $i \neq j \neq k$.

In momentum representation the operators $X$ and $Z$ are integrated over
the Jacobi momenta $\vec{k_{i}}^{\alpha}$. Therefore, 
the operators act on the second momentum, $\vec{p_{i}^{\alpha}}$
\begin{eqnarray}
\label{X_mom}
&{}& 
\left\langle \vec{p_i}^{\alpha} | X_{ij, I_i I_j}^{\alpha \beta}
(z_{tot}) | \vec{p_j}'^{\beta} \right \rangle
=  X_{ij, I_i I_j}^{\alpha \beta}
(\vec{p_i}^{\alpha}, \vec{p_j}'^{\beta}; z_{kin}^{\alpha} + z_{th}^{\alpha}) \,, \\
\label{Z_mom}
&{}&
\left\langle \vec{p_i}^{\alpha} | Z_{ij, I_i I_j}^{\alpha}
(z_{tot}) |
\vec{p_j}'^{\alpha} \right \rangle
=  Z_{ij, I_i I_j}^{\alpha}
(\vec{p_i}^{\alpha}, \vec{p_j}'^{\alpha}; z_{kin}^{\alpha} + z_{th}^{\alpha})
\,,
\end{eqnarray}
where the total energy  $z_{tot} = z_{th}^{\alpha} + z_{kin}^{\alpha}$
is the sum of the channel kinetic energy  $z_{kin}^{\alpha}$ and the
channel threshold energy $z_{th}^{\alpha} = \sum_{i=1}^3 m_i^{\alpha}$.
The energy-dependent part of a two-body $T$-matrix, embedded in the
three-body space, is defined by the following relation:
\begin{equation}
\label{tau_mom}
\left\langle \vec{p_i}^{\alpha} | \tau_{i, I_i}^{\alpha \beta}(z_{tot}) |
\vec{p_j}'^{\beta} \right \rangle  \equiv
\delta_{ij} \, \delta(\vec{p_i}^{\alpha} - \vec{p_j}'^{\beta}) \,
\tau_{i,I_i}^{\alpha \beta}\left(
z_{tot} - z_{th}^{\alpha} - \frac{(p_i^{\alpha})^2}{2 \, \mu_i}
\right) \,.
\end{equation}

Since the antikaon is a pseudoscalar meson, the total spin of the
$\bar{K} \bar{K} N$ system is equal to one half. All our two-body
interactions have zero orbital angular momentum, therefore, the total
angular momentum $J$ is also $1/2$. As for the isospin, we consider the 
three-body system with the lowest possible value, i.e. with $I^{(3)}=1/2$.

In what follows the indices on the right-hand side of the operators
$X_{ij, I_i I_j}^{\alpha \beta}$ will be omitted, i.e. 
$X_{ij, I_i I_j}^{\alpha \beta} \to X_{i, I_i}^{\alpha}$,
since they denote the initial state, which is fixed for a given system.
To search for a quasi-bound state means to look for a solution of
the homogeneous equations 
\begin{equation}
\label{X_hom}
 X_{i,I_i}^{\alpha} = \sum_{k=1}^3 \sum_{\gamma=1}^3
  \sum_{I_k} Z_{ik,I_i I_k}^{\alpha} \, \tau_{k,I_k}^{\alpha \gamma}
   X_{k,I_k}^{\gamma},
\end{equation}
which however, should be symmetrized first since there are two identical
mesons in the $\bar{K}\bar{K}N$ system. The $X_{3,1}^1$ transition operator
has the proper symmetry properties already (while $X_{3,0}^1$ is antisymmetric).
The remaining operators acquire the necessary symmetry properties in
the following combinations:
\begin{eqnarray}
\nonumber
&{}&X_{1,0}^{1,sm} = X_{1,0}^{1} + X_{2,0}^{1}, \quad
 X_{1,1}^{1,sm} = X_{1,1}^{1} + X_{2,1}^{1}, \\
\nonumber
&{}&X_{1,0}^{2,sm} = X_{1,0}^{2} + X_{2,0}^{3}, \quad
 X_{1,1}^{2,sm} = X_{1,1}^{2} + X_{2,1}^{3}, \\
\label{X_symmetrical}
&{}&X_{2,\frac{1}{2}}^{2,sm} = X_{2,\frac{1}{2}}^{2} + X_{1,\frac{1}{2}}^{3},
\quad
 X_{2,\frac{3}{2}}^{2,sm} = X_{2,\frac{3}{2}}^{2} + X_{1,\frac{3}{2}}^{3}, \\
\nonumber
&{}&X_{3,\frac{1}{2}}^{2,sm} = X_{3,\frac{1}{2}}^{2} - X_{3,\frac{1}{2}}^{3},
\quad
 X_{3,\frac{3}{2}}^{2,sm} = X_{3,\frac{3}{2}}^{2} + X_{3,\frac{3}{2}}^{3} \,.
\end{eqnarray}

In momentum representation the system of operator equations (\ref{X_hom})
turns into a system of integral equations, schematically given by 
\begin{equation}
\label{AGS_imp}
 X_{i}(p) = \int_{0}^{\infty} Z_{ij}(p,p';z) \,
 \tau_{j}\left(z - \frac{p'^2}{2 \mu_j}\right) X_{j}(p') dp'.
\end{equation}
This system is then discretized, and the value of $z$ at which the determinant of the 
kernel matrix $A_{mn}(z) = [Z(z) \, \tau(z)]_{mn}$ is equal to zero is determined. 
This complex energy, located on the proper energy sheet, corresponds to 
the pole of the quasi-bound state.

In the present study we determine the pole position in two ways. First, we 
perform a direct search of the pole in the complex energy plane as described 
above. For that purpose we have to rotate the contour of the integration in 
Eq.~(\ref{AGS_imp}) into the complex momentum plane in order to avoid
irregular regions and singularities, see Ref.~\cite{PRC_II} for a detailed discussion. 

Since the analytic continuation of the integral equations into the complex plane is a 
non-trivial procedure, we employ here in addition the inverse determinant method
proposed and successfully applied in Ref.~\cite{PRC_II} for the $K^- pp$ system. 
It uses the fact that the function $1/|Det\, A(z)|^2$, calculated along the real energy 
axis $z$, exhibits a resonance shape in the neighbourhood of the pole           
position. This bump can be fitted using the Breit-Wigner type formula, and
the position and the width of the resonance can be estimated. Since the 
calculations in this case are performed on the real energy axis, it allows to avoid
possible problems with the analytic continuation of the equations into
the complex momentum plane.

It was demonstrated in Ref.~\cite{PRC_II} that the two methods are indeed
complementary. However, it is necessary to keep in mind that the second method
gives a reliable estimation of the pole position only in the case where 
the resonance is quite narrow and, therefore, the produced bump is
fairly pronounced.

\section{Interactions}
\label{Interactions_sect}

The explict form of the one-term separable potentials, introduced
in Eq.~(\ref{Voperator}), in momentum representation is
\begin{equation}
  V_{i,I_i}^{\alpha \beta}(k_i^{\alpha},{k'}_i^{\beta}) = 
   \lambda_{i,I_i}^{\alpha \beta} \, g(k_i^{\alpha}) g({k'}_i^{\beta}).
\label{VSep}
\end{equation}

The used $\bar{K}N$ and $\bar{K}\bar{K}$ interactions are described in the
following subsections. The remaining interactions in the three-body system
with coupled $\bar{K}\bar{K}N$ and $\bar{K} \pi \Sigma$ channels are
those in the lower-lying three-body channel, namely $\pi \Sigma$,
$\bar{K}\Sigma$, and $\bar{K}\pi$.  
The first one, $\pi \Sigma$, is part of the employed coupled-channel model 
for the $\bar{K}N$ interaction. Almost nothing is known about the $\bar{K} \Sigma$
interaction, in particular, there is no experimental information. There
are suggestions~\cite{Ramos:2002,Sekihara:2015} that  
this strangeness $S=-2$ system can form and couple to $\Xi$ resonances.
The $\bar{K} \pi$ system is related to $K \pi$ via charge conjugation
and for the latter phase shifts are available  
\cite{Aston:1987,Linglin:1973,Estabrooks:1977}.
Thus, in principle, it would be possible to construct a potential 
in a similar way as for the $\bar{K}N$ interaction described below
by fitting its parameters to those phase shifts.
However, we assume that these two interactions in the lower 
three-body channel are not so important for the system under consideration
and can be omitted. In any case, keeping in mind the unclear situation with
regard to the $\bar{K} \Sigma$ interaction, an inclusion of those channels into 
the calculation would not lead to quantitatively better constrained
$\bar{K}\bar{K}N$ quasi-bound state results.

Therefore, our three-body calculation with the coupled $\bar{K}\bar{K}N-
\bar{K} \pi \Sigma$ channels has only one non-zero interaction
in the lower channel: the $\pi \Sigma$, coupled to the $\bar{K}N$.
Then our two-channel three-body calculation with coupled-channel
$\bar{K}N-\pi \Sigma$ potential is equivalent to the one-channel
three-body calculation utilizing the so-called exact optical $\bar{K}N$ potential,
see \cite{my_Kd_PRC}. We performed calculations based on both formulations and
obtained perfect agreement between their results.

\subsection{Antikaon-nucleon interaction}
\label{KNint_sect}

Several models of the $\bar{K}N$ interaction were constructed by us in the past 
for application in our works devoted to the $\bar{K}NN$ system. In the present
study we employ three of those.  In particular, we use the two phenomenological
$\bar{K}N - \pi \Sigma$ potentials that yield a one- or two-pole structure
of the $\Lambda(1405)$ resonance which were presented in~\cite{my_Kd_sdvig}.
The form factors of the one-pole version of the potential and
those of the $\bar{K}N$ channel of the two-pole version have
a Yamaguchi form
\begin{equation}
 g_I^{\alpha} = \frac{1}{(k^{\alpha})^2 + (\beta_I^{\alpha})^2},
\end{equation}
while a slightly more complicated form is used for the $\pi \Sigma$
channel 
\begin{equation}
 g_I^{\alpha} = \frac{1}{(k^{\alpha})^2 + (\beta_I^{\alpha})^2} +
  \frac{s \, (\beta_I^{\alpha})^2}
        {\left[ (k^{\alpha})^2 + (\beta_I^{\alpha})^2 \right]^2}
\end{equation}
in the two-pole model of the interation.

Recently, a chirally motivated potential describing the coupled
$\bar{K}N - \pi \Sigma - \pi \Lambda$ channels was constructed and used
in~\cite{PRC_I}. In contrast to the energy independent phenomenological 
models mentioned above the chirally motivated
potential has strength constants $\lambda_{i,I_i}^{\alpha \beta}(z^{(2)})$,
which depend on the energy in the two-body subsystem $z^{(2)}$.

All three potentials were fitted to data on $\bar{K}N$ scattering and 
characteristics of kaonic hydrogen. In particular, the potentials reproduce 
the measured cross sections of elastic 
($K^- p \to K^- p$) and inelastic ($K^- p \to \bar{K}^0 n$,
$K^- p \to \pi^+ \Sigma^-$, $K^- p \to \pi^- \Sigma^+$,
$K^- p \to \pi^0 \Sigma^0$, $K^- p \to \pi^0 \Lambda$)
scattering (the last reaction is described by the chirally motivated 
potential only) from different
experiments~\cite{Kp2exp,Kp3exp,Kp4exp,Kp5exp,Kp6exp}.

They also reproduce the accurately measured threshold branching ratios
$\gamma$, $R_c$ and $R_n$~\cite{gammaKp1, gammaKp2}
\begin{eqnarray}
\label{gamma}
\gamma &=& \frac{\Gamma(K^- p \to \pi^+ \Sigma^-)}{\Gamma(K^- p \to
\pi^- \Sigma^+)} = 2.36 \pm 0.04  \\
\label{Rc}
R_c &=& \frac{\Gamma(K^- p \to \pi^+ \Sigma^-, \pi^- \Sigma^+)}{\Gamma(K^- p \to
\mbox{all inelastic channels} )} = 0.664 \pm 0.011, \\
\label{Rn}
R_n &=& \frac{\Gamma(K^- p \to \pi^0 \Lambda)}{\Gamma(K^- p \to
\mbox{neutral states} )} = 0.189 \pm 0.015.
\end{eqnarray}
Since the $\pi \Lambda$ channel is taken into account in the phenomenological
potentials indirectly, through an imaginary part of one of the
$\lambda^{\alpha \beta}$ parameters, we constructed a new threshold branching
ratio based on $R_c$ and $R_n$:
\begin{equation}
 R_{\pi \Sigma} =
 \frac{\Gamma(K^- p \to \pi^+ \Sigma^-)+\Gamma(K^- p \to \pi^- \Sigma^+)}{
 \Gamma(K^- p \to \pi^+ \Sigma^-) + \Gamma(K^- p \to \pi^- \Sigma^+) +
                                   \Gamma(K^- p \to \pi^0 \Sigma^0) } \,,
\end{equation}
which has an ``experimental'' value (derived from the experimental data on
$R_c$ and $R_n$)
\begin{equation}
 R_{\pi \Sigma} =  \frac{R_c}{1-R_n \, (1 - R_c)} \, = \, 0.709 \pm 0.011 \,.
\end{equation}
The chirally motivated potential reproduces all three experimental
branching ratios directly.

Finally, all three $\bar{K}N$ models reproduce the most
recent experimental results of the SIDDHARTA experiment~\cite{SIDDHARTA} 
on the $1s$
level shift $\Delta E_{1s}$ and width $\Gamma_{1s}$ of kaonic hydrogen
\begin{equation}
\Delta E_{1s} = -283 \pm 36 \pm 6 \; {\rm eV}, \quad
\Gamma_{1s} = 541 \pm 89 \pm 22 \; {\rm eV}.
\end{equation}
Note that those quantities were calculated directly, without using some
approximate formula.
All $\bar{K}N$ results were obtained
by solving coupled-channels Lippmann-Schwinger equation with the strong
interaction plus the Coulomb potential. In addition, the physical masses
of the involved particles were used, so that the associated two-body
isospin nonconservation is properly included. However, the three-body
calculations are performed with isospin averaged masses and without
Coulomb interaction for simplicity reasons but also because we expect
the pertinent effects to be small.

Irrespective of the number of poles that constitute the $\Lambda(1405)$
resonance, which appears as a quasi-bound state in the $\bar{K}N$
channel and as a resonance in the lower channels, for all considered potentials
the resulting isospin-zero elastic $\pi \Sigma$ cross sections has a peak
near the position of the resonance as given by the Particle Data Group (PDG) 
with comparable width ($M_{\Lambda(1405)} = 1405.1$ MeV,
$\Gamma_{\Lambda(1405)} = 50.5$ MeV according to the most recent issue~\cite{PDG}). 

All three $\bar{K}N$ potentials describe the experimental information 
with the same level of accuracy, as one can see in Refs.~\cite{my_Kd_sdvig}
and \cite{PRC_I} for the phenomenological and chirally motivated
potentials, respectively. The actual parameters of the potentials 
can be found in those papers.

\subsection{Antikaon-antikaon interaction}
\label{KKint_sect}

There is no experimental information on the $\kbarkbar$ interaction and,
therefore, the $\bar{K}\bar{K}$ potential cannot be constructed in the same
way as the one for $\bar{K} N$. Hence, we resort to theory and adopt here
the  $\pi\pi - \kkbar$ coupled channels model developed by the 
J\"ulich group \cite{Lohse90,Janssen95} some time ago as guideline. 
Indeed, based on the underlying SU(3) flavor symmetry the interaction in 
the $\kbarkbar$ system (or equivalently in the $KK$ system) can be directly 
deduced from the $\kkbar$ interaction of Ref.~\cite{Janssen95} without any
further assumptions. 
A detailed description of the J\"ulich $\pi\pi - \kkbar$ meson exchange model
can be found in Refs.~\cite{Lohse90,Janssen95}. Here we provide only a short
summary of the interaction. 

\begin{figure}[t]
\includegraphics[width=0.90\textwidth]{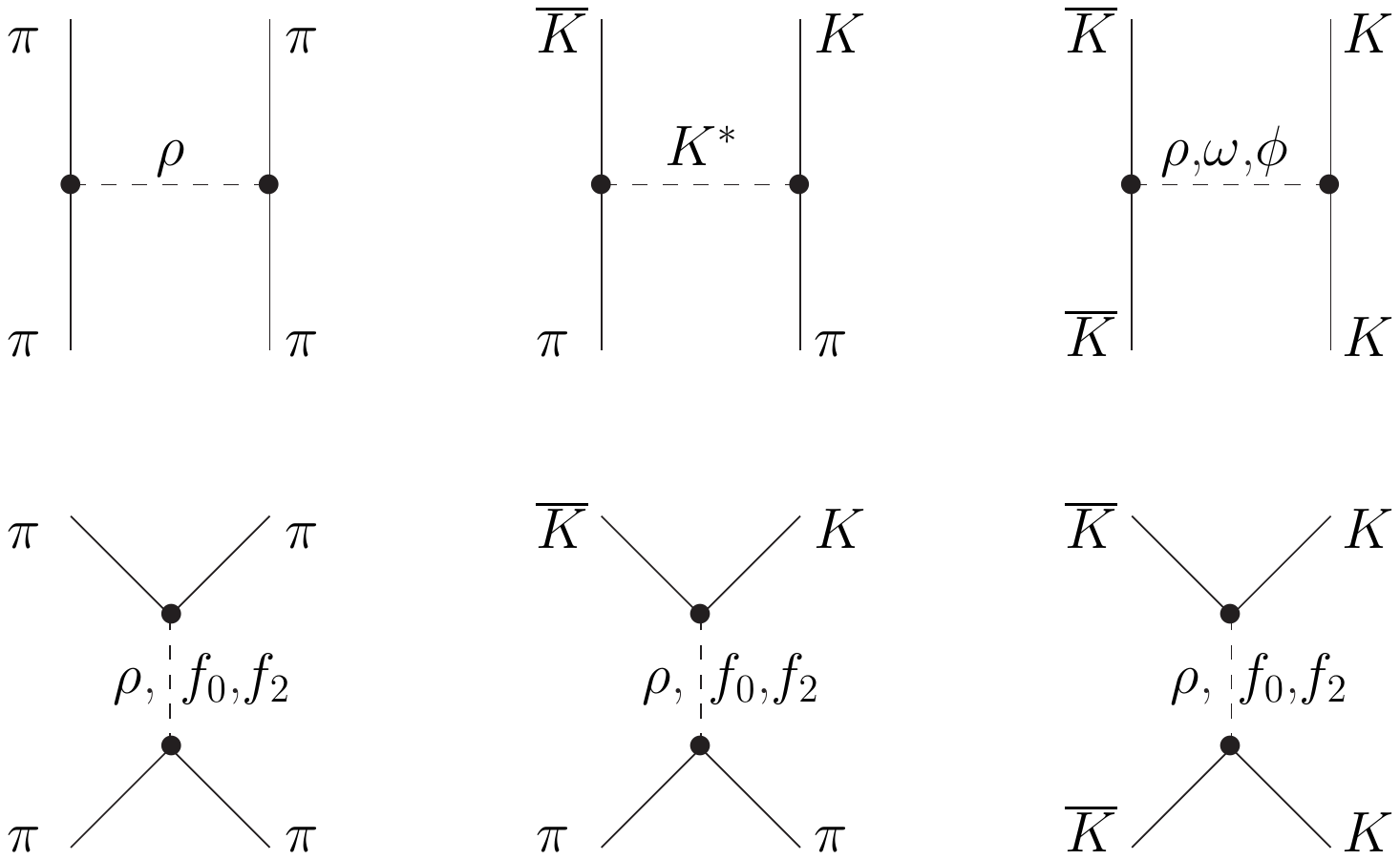}
\caption{Diagrams included in the J\"ulich
$\pi\pi - \kkbar$ potential~\cite{Janssen95}.
}
\label{Diag1}
\end{figure}

The dynamical input that constitutes the J\"ulich $\pi\pi - \kkbar$ model 
is depicted in Fig.~\ref{Diag1}. The figure contains only $s$- and $t$-channel 
diagrams; $u$-channel processes corresponding to the considered $t$-channel 
processes are also included
whenever they contribute. The scalar-isoscalar particle denoted by
$f_0$ in Fig.~\ref{Diag1} (and $\epsilon$ in Ref.~\cite{Janssen95})
effectively includes the singlet and the octet member of the scalar nonet.  
The effects of $t$-channel $f_2(1270)$ and $f_0$ exchange were 
found to be negligible \cite{Janssen95} and, therefore, not
included in the model. 

The coupling constant $g_{\rho\pi\pi}$, required for $t$- and
$u$-channel exchange diagrams, is determined from the decay widths of
the $\rho$. Most of the other coupling constants are determined from 
SU(3) symmetry relations, and standard assumptions about the 
octet/singlet mixing angles, as described in Ref.~\cite{Lohse90}.
The J\"ulich $\pi\pi - \kkbar$ potential contains also 
vertex form factors and those are parametrized in the conventional 
monopole or dipole form, cf. the Appendix of Ref.~\cite{Janssen95}. 
The values of the inherent cutoff masses have been determined in a 
fit to the $\pi\pi$ phase shifts.

This interaction yields a good description of the $\pi\pi$
phase shifts up to partial waves with total angular momentum
$J=2$ and for energies up to $z_{\pi\pi}\approx$ 1.4 GeV as
can be seen in Ref.~\cite{Janssen95}. Furthermore, as a special
feature, the $f_0(980)$ meson results as a dynamically generated 
state, namely as a quasi-bound $\bar KK$ state. 
Also the $a_0(980)$ is found to be dynamically generated in
the corresponding $\pi\eta-\bar KK$ system. 

The interaction in the $\kbarkbar$ (or the $KK$) system follows directly
from the one for $\kkbar$ by invoking SU(3) symmetry arguments. It is 
provided by vector-meson exchange
($\rho$, $\omega$, $\phi$) with coupling constants fixed according
to standard SU(3) relations, see Table I of \cite{Janssen95}. For identical
particles the Bose-Einstein statistics applies and it restricts 
the $\kbarkbar$ $s$-wave to be solely in isospin $I=1$. In this 
case the contributions of the three vector-meson exchanges add up 
coherently and they are all repulsive so that one expects an
overall repulsive interaction in the $\kbarkbar$ $s$-wave.
Indeed the $\kbarkbar$ scattering length predicted by the J\"ulich
model is $a_{\bar{K}\bar{K},I=1}= -0.186$ fm. This version of
the $\bar{K}\bar{K}$ interaction will be called 'Original'.

Recently, results for the $KK$ scattering length have become available
from lattice QCD simulations \cite{Sasaki:2013vxa,Beane:2007uh}.
Those calculations, performed for quark masses corresponding to 
$m_\pi = 170-710$ MeV, suggest values of
$a_{\bar{K}\bar{K},I=1}= (-0.141\pm 0.006)$ fm \cite{Beane:2007uh} and
$a_{\bar{K}\bar{K},I=1}= (-0.124\pm 0.006\pm 0.013)$ fm \cite{Sasaki:2013vxa}, 
respectively, when extrapolated to the physical point. 
Since those values are noticeably smaller than
the one predicted by the J\"ulich meson-exchange model and, 
accordingly, imply a somewhat less repulsive $\kbarkbar$
interaction we construct also an interaction that is in line
with the lattice QCD results. A corresponding potential can be easily
generated by simply reducing the values of the cutoff masses
for the vector-meson exchange in the J\"ulich model until the scattering 
length suggested
by the lattice QCD calculations is reproduced. The interaction
constructed with that aim yields $a_{\bar{K}\bar{K},I=1}= -0.142$ fm,
close to the result by the NPLQCD collaboration \cite{Beane:2007uh}.
This version of the $\bar{K}\bar{K}$ interaction will be called
'Lattice motivated'.

\begin{figure}
\includegraphics[width=0.80\textwidth]{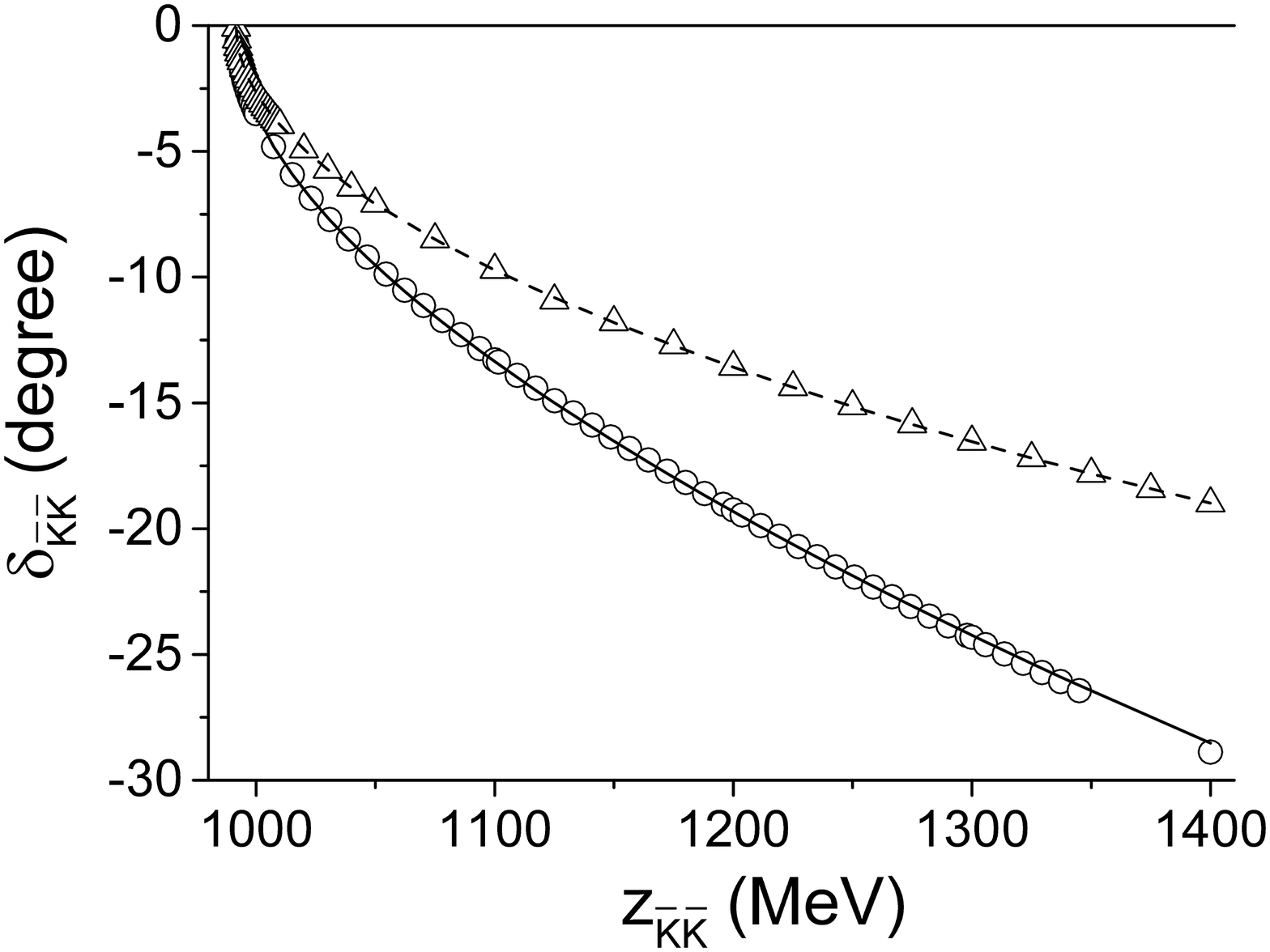}
\caption{$\kbarkbar$ $s$-wave phase shifts in the $I=1$ state.
The circles and
the solid line denote the result of the ``Original'' J\"ulich model
and its separable representation, respectively. 
The triangles and the dashed line are corresponding results for 
the ``Lattice motivated'' $\bar{K}\bar{K}$ potentials.
}
\label{phases}
\end{figure}

We cannot use the models of the $\bar{K}\bar{K}$ interaction 
described above directly in the AGS equations. Therefore, we
represent also the $\bar{K}\bar{K}$ interaction in form of one-term separable 
potentials, see Eq.~(\ref{VSep}), with form factors given by 
\begin{equation}
 g(k) = \frac{1}{\beta_1^2 + k^2} +
  \frac{\gamma}{\beta_2^2 + k^2}.
\end{equation}
The strength parameters $\lambda$, $\gamma$ and range parameters
$\beta$ are fixed by a fit to the $\bar{K}\bar{K}$ phase shifts and
scattering lengths of the 'Original' J\"ulich model and the 'Lattice motivated' 
interaction. The phase shifts predicted by the initial interactions and those 
of the corresponding separable potentials are displayed in Fig.~\ref{phases},
so that one can see the quality of the reproduction.

\section{Results and discussion}
\label{Results_sect}

The results of our calculations are summarized in Table~\ref{Pole1Det_KK.tab} 
for various combinations of the employed $\bar{K}\bar{K}$ and $\bar{K}N$ 
interactions. We used the two $\bar{K}\bar{K}$ interactions 
(Original $V_{\bar{K}\bar{K}}^{Orig}$ and
Lattice-motivated $V_{\bar{K}\bar{K}}^{Latt}$) described in the Section III B
and three $\bar{K}N$ potentials, namely the phenomenogical
one-pole $V^{1,SIDD}_{\bar{K}N - \pi \Sigma}$ and
two-pole $V^{2,SIDD}_{\bar{K}N - \pi \Sigma}$ potentials 
from~\cite{my_Kd_sdvig}, and the chirally-motivated
$V^{Chiral}_{\bar{K}N - \pi \Sigma - \pi \Lambda}$ from~\cite{PRC_I}
discussed in Section~\ref{KNint_sect}.
The pole positions of the quasi-bound state in the $\bar{K}\bar{K}N$ system
were determined by a direct search in the complex momentum plane and by using
the inverse determinant method~\cite{PRC_II}. 
\begin{center}
\begin{table*}[ht]
\caption{Pole positions (in MeV) of the quasi-bound
state in the $\bar{K}\bar{K}N$ system. Results of the direct pole search and
of the inverse determinant method are given, employing various combinations
of the $\bar{K}\bar{K}$ and $\bar{K}N$ in the AGS equations, see description
in the text. Two-body pole position(s) of the $\bar{K}N$ potentials are also
presented. The two- and three-body thresholds are situated at $1434.57$ and
$1930.21$ MeV, respectively.}
\label{Pole1Det_KK.tab}
\begin{center}
\begin{tabular}{ccccc}
\hline \noalign{\smallskip}
  & 
  &  $\bar{K}\bar{K}N$ pole, &  $\bar{K}\bar{K}N$ pole, & $\bar{K}N$ pole(s) \\
  & 
  &  direct search &  \ inverse determinant \ &  \\
       \noalign{\smallskip} \hline \noalign{\smallskip}
 &  $V^{1,SIDD}_{\bar{K}N - \pi \Sigma}$ 
           & $1918.31 -  i \, 51.14$ & $1913.14 -  i \, 55.39$ & 1428.14 - i 46.81\\
$V_{\bar{K}\bar{K}}^{Orig}$ 
 & $V^{2,SIDD}_{\bar{K}N - \pi \Sigma}$  
           & $1907.15 -  i \, 45.69$ &  $1906.49 -  i \, 38.81$ & 1418.11 - i 57.01\\
     &  & & & 1382.03 - i 104.15\\
 & $V_{\bar{K}N - \pi \Sigma - \pi \Lambda}^{\rm Chiral}$ 
           & $1914.70 -  i \, 31.75$ & $1914.34 -  i \, 28.71$ & 1418.08 - i 32.83\\
     &  & & & 1407.03 - i 88.31\\
           \noalign{\smallskip} \hline \noalign{\smallskip}
 
 & $V^{1,SIDD}_{\bar{K}N - \pi \Sigma}$ 
           & $1910.70 -  i \, 51.01$ & $1906.51 -  i \, 51.85$ & 1428.14 - i 46.81\\
$V_{\bar{K}\bar{K}}^{Latt}$ 
 & $V^{2,SIDD}_{\bar{K}N - \pi \Sigma}$ 
           & $1904.28 -  i \, 42.30$ &  $1903.81 -  i \, 38.39$ & 1418.11 - i 57.01\\
     &  & & & 1382.03 - i 104.15\\
 & $V_{\bar{K}N - \pi \Sigma - \pi \Lambda}^{\rm Chiral}$ 
           & $1914.12 -  i \, 30.66$ & $1914.27 -  i \, 29.96$ & 1418.08 - i 32.83\\
     &  & & & 1407.03 - i 88.31\\
           \noalign{\smallskip} \hline             
\end{tabular}
\end{center}
\end{table*}
\end{center}
 
It is seen from the Table that all combinations of the two-body interactions
lead to a quasi-bound state in the $\bar{K}\bar{K}N$ three-body system.
Thus, the repulsion in the $\bar{K}\bar{K}$ subsystem is more than
compensated by the attraction provided by the interaction in the 
$\bar{K}N$ subsystem(s) and does not prevent the formation of a quasi-bound state. 

Comparing the poles for $\bar{K}\bar{K}N$, obtained with the two-pole models of
the $\bar{K}N$ interaction, we see that the chirally motivated
potential leads to a more shallow quasi-bound state than the phenomenological
one. We believe that this must be connected with the
energy dependence of the former potential, because 
the real parts of the higher two-body pole position corresponding
to the $\Lambda(1405)$ resonance are almost the same for the both potentials,
see Table~\ref{Pole1Det_KK.tab} (right column). Comparing the widths
of the states we see that for the 
chirally motivated $V^{Chiral}_{\bar{K}N}$ the width of the 
$\bar{K}N$ state and that of the $\bar{K}\bar{K}N$ state are almost the same.
The phenomenological $V^{2,SIDD}_{\bar{K}N}$ with a 
broader ``$\Lambda(1405)$''  leads to a noticeably more narrow $\bar{K}\bar{K}N$
state, which, however, is still wider than the ''chirally motivated" one.
The situation is opposite for the one-pole $V^{1,SIDD}_{\bar{K}N}$
model, where the two-body quasi-bound $\bar{K}N$ state is more
narrow than the three-body $\bar{K}\bar{K}N$ state.
In this context we want to mention that a somewhat surprising behaviour
of the results was also observed in~\cite{JanosKd} in the
$K^- d \to \pi \Sigma N$ spectra based on the $V^{1,SIDD}_{\bar{K}N}$ potential
-- in contrast to the ones for other phenomenological ${\bar{K}N}$ models.
Since another one-pole potential used in~\cite{JanosKd} behaves quite normaly,
it seems that the different trend seen in the calculations with
$V^{1,SIDD}_{\bar{K}N}$ could be due to some peculiar features of this particular
one-pole potential.

The results obtained with the phenomenological $\bar{K}N$ potentials 
exhibit more sensitivity to the $\bar{K}\bar{K}$ interaction than
the chirally motivated interaction. As was expected, the less
repulsive $\bar{K}\bar{K}$ model that simulates the lattice QCD results
leads to somewhat deeper quasi-bound three-body state than the 
$\bar{K}\bar{K}$ interaction based on the original J\"ulich
$\pi\pi - \bar KK$ model. The largest difference between results based on
the potentials $V_{\bar{K}\bar{K}}^{Orig}$ and $V_{\bar{K}\bar{K}}^{Latt}$
were obtained with the one-pole phenomenological $V^{1,SIDD}_{\bar{K}N}$
model. A less repulsive $\bar{K}\bar{K}$ interaction
also leads to a more narrow $\bar{K}\bar{K}N$ quasi-bound state in all the
cases. 

A comparison of the pole positions obtained from the direct pole search
to the ones that follow from the inverse determinant method reveals 
that the accuracy of the second method is much lower for the phenomenological
$\bar{K}N$ interactions than for the chirally motivated one. This is
an expected result keeping in mind the larger widths of the ''phenomenogical''
$\bar{K}\bar{K}N$ states. The one-to-one connection between
a complex pole and the Breit-Wigner form of the corresponding bump on the real
axis is obviously less pronounced if the pole is situated further away from
the real axis.

It is also interesting to compare the binding energies and widths of the quasi-bound 
states in the $\bar{K}\bar{K}N$ system with those for $\bar{K}NN$,
obtained in~\cite{PRC_II}.
In both three-body systems the strongly attractive $\bar{K}N$ interaction
is present and plays an essential role. We see that for a specific
$\bar{K}N$ potential the quasi-bound state in the strangeness $S=-2$
$\bar{K}\bar{K}N$ system is only about half as deep as that in the
$S=-1$ three-body system. The $\bar{K}\bar{K}N$ states are
also much broader, especially those obtained with the phenomenological
$\bar{K}N$ models. The differences in the binding energies are expected
since the $NN$ interaction appearing in $\bar{K}NN$ is attractive, while
the $\bar{K}\bar{K}$ interaction in $\bar{K}\bar{K}N$ is repulsive.
Their (attractive or repulsive) character is, probably, the origin of
the differences in the widths too.

As already said in the Introduction, there has been a previous investigation
on the $\bar{K}\bar{K}N$ system~\cite{KanadaEn'yo:2008wm}. Though the binding 
energies reported in that work are of the same order of magnitude
we want to emphasize that in reality it is difficult to compare our results
with the ones in that paper. The authors of~\cite{KanadaEn'yo:2008wm}
used energy-independent as well as 
energy-dependent potentials, but the two-body energy of the latter is fixed
arbitrarily. Moreover, the imaginary parts of all complex potentials
are completely ignored in their variational calculations. That imaginary
parts are treated only perturbatively, which, probably, is one of the
reasons of the strong underestimation of the $\bar{K}\bar{K}N$ widths
in comparison to ours. In a series of works devoted to the $\bar{K}NN$ 
system we demonstrated that a proper inclusion of the lower-lying channels -- 
either by a direct inclusion or by using the 
exact optical potential, see e.g.~\cite{our_KNN_PRC,my_Kd_PRC} 
-- is very important. Therefore, the results of Ref.~\cite{KanadaEn'yo:2008wm}, 
involving several uncontrolled approximations, can be considered only 
as a very rough estimation.

\section{Conclusions}
\label{Conclusion_sect}

We presented a dynamically exact calculation of a quasi-bound state in
the $\bar{K}\bar{K}N$ three-body system, performed in the framework of
Faddeev-type AGS equations. As input we used two phenomenological and one
chirally motivated $\bar{K}N$ potentials, which describe the experimental information
on the $\bar{K}N$ system equally well and produce either a one- or two-pole structure
of the $\Lambda(1405)$ resonance. For the $\bar{K}\bar{K}$ subsystem we resort to an
interaction
that is adjusted to the $KK$ $s$-wave scattering length recently determined in
lattice QCD simulations. In addition a phenomenological $\bar{K}\bar{K}$ potential
based on meson-exchange is employed, which is derived by SU(3) symmetry arguments
from the J\"ulich  $\pi\pi - \bar{K}K$ coupled-channels model.
The position and width of the $\bar{K}\bar{K}N$ quasi-bound state were evaluated
in two ways: (i) by a direct pole search in the complex energy plane and 
(ii) using the inverse determinant method.

We found that a quasi-bound state exists in the $\bar{K}\bar{K}N$
system in spite of the repulsive character of the $\bar{K}\bar{K}$
interaction. Its binding energy and width are in the region
$12 - 26$ MeV and $61 - 102$ MeV, respectively, where the variation
reflects 
the uncertainty due to differences in the two-body input.
The quasi-bound state in the strangeness $S=-2$ $\bar{K}\bar{K}N$ system turned out
to be much shallower and broader than the one in the $S=-1$ $\bar{K}NN$ system,
when comparing calculations with the same $\bar{K}N$ potential.

What are the perspectives of finding the $\bar{K}\bar{K}N$ quasi-bound
state in experiments? This state has the same quantum numbers as a $\Xi$ baryon
with $J^P=(1/2)^+$, as already noted in~\cite{KanadaEn'yo:2008wm}.
The available experimental information on the $\Xi$ spectrum is rather
limited, see \cite{PDG}, and for $(1/2)^+$ only the ground state (with an 
isospin-averaged mass of 1318 MeV) 
is established. The quark model \cite{Isgur} predicts a first excited state 
at around 1700 MeV and another one around 1950 MeV, where the latter is
already above the $\bar{K}\bar{K}N$ threshold. There is a $\Xi$(1950) listed
by the PDG but its quantum numbers $J^P$ are not determined and it is 
unclear whether it should be identified with the quark-model state. Indeed the 
PDG suggests that there could be even more than one resonance in this region
because the mass values of the experimental evidence summarized in the
listing \cite{PDG} scatter over 70 MeV or so. It is interesting to see that
four of the values would be roughly consistent with the quasi-bound state found
in the present study. Specifically, the experiment reported in
Ref.~\cite{Dauber:1969} yielded a mass of $1894 \pm 18$ MeV and a width of $98 \pm 23$
MeV that is more or less compatible with the range of values for the pole
position that emerged from our three-body calculation. 
In any case, in view of concrete plans for experiments dedicated to $\Xi$ 
baryon spectroscopy at J-PARC \cite{JPARC} and JLAB \cite{JLABprev,JLAB}
there are good chances that more information about the strangeness $S=-2$
resonances will become available. Probably, then one can draw
more solid conclusions on the $\bar{K}\bar{K}N$ quasi-bound state.

\vspace{5mm}

\noindent
{\bf Acknowledgments.}
The work was supported by the Czech CACR grant 15-04301S. One of the authors
(NVS) is thankful to J. R\'evai for fruitful discussions.

\end{document}